\documentclass[aps,jcp,preprint,a4paper,onecolumn]{revtex4}

\usepackage[fleqn]{amsmath}
\usepackage{amssymb}
\usepackage[dvips]{graphicx}
\usepackage{color}
\usepackage{tabularx}
\usepackage{algpseudocode}


\newcommand{\redc}[1]{{#1}}

\newcommand{\vect}[1]{#1}
\newcommand{\mypi}{\pi}

\newcommand{\haml}[0]{\mathcal {H}}
\newcommand{\opfwd}     {\mathcal F}

\newcommand{\dvs}{\dot{\vect s}}
\newcommand{\dvh}{\dot{\vect h}}
\newcommand{\dvr}{\dot{\vect r}}
\newcommand{\dvp}{\dot{\vect p}}
\newcommand{\ins}{\textrm{ins}}

\newcommand{\equi}{\mathrm{equi}}

\newcommand{\tensorab}{{\alpha\beta}}
\newcommand{\tensoraa}{{\alpha\alpha}}
\newcommand{\inv}{{-1}}
\newcommand{\phys}{\mathrm{phys}}
\newcommand{\kt}{{k_{_\textrm{B}}T}}

\begin{document}

\title{Sampling the isothermal-isobaric ensemble by Langevin dynamics}
\author{Xingyu Gao}
\affiliation{Laboratory of Computational Physics, Huayuan Road 6, Beijing 100088, P.R.~China}
\affiliation{Institute of Applied Physics and Computational Mathematics, Fenghao East Road 2, Beijing 100094, P.R.~China}
\affiliation{CAEP Software Center for High Performance Numerical Simulation, Huayuan Road 6, Beijing 100088, P.R.~China}
\author{Jun Fang}
\affiliation{Institute of Applied Physics and Computational Mathematics, Fenghao East Road 2, Beijing 100094, P.R.~China}
\affiliation{CAEP Software Center for High Performance Numerical Simulation, Huayuan Road 6, Beijing 100088, P.R.~China}
\author{Han Wang}
\email{wang_han@iapcm.ac.cn}
\affiliation{Institute of Applied Physics and Computational Mathematics, Fenghao East Road 2, Beijing 100094, P.R.~China}
\affiliation{CAEP Software Center for High Performance Numerical Simulation, Huayuan Road 6, Beijing 100088, P.R.~China}
   
\begin{abstract}
  We present a new method of conducting fully-flexible-cell molecular dynamics simulation in isothermal-isobaric ensemble
  based on Langevin equations of motion.
  The stochastic coupling to all particle and cell degrees of freedoms is introduced in a correct way,
  in the sense that the stationary configurational distribution
  is proved to be in consistent with that of the isothermal-isobaric ensemble.
  In order to apply the proposed method in computer simulations,
  a second order symmetric numerical integration scheme
  is developed by Trotter's splitting of the single-step propagator.  
  Moreover, a practical guide of choosing working parameters is suggested for user specified thermo- and baro-coupling time-scales.
  The method and software implementation are carefully validated by a numerical example.
\end{abstract}

\maketitle

\section{Introduction}
\label{sec:intro}

Molecular dynamics (MD) simulation is a powerful tool for investigating a broad range of systems, from biological to materials sciences.
In the equilibrium situation, it is of crucial importance to consider the ensemble that an MD simulation samples,
because the quantities to observe are often calculated from the ensemble averages.
Moreover,
the equilibrium ensemble may serve as the initial condition for non-equilibrium MD simulations~\cite{wang2014exploring}.
Most of early MD simulations solve the Hamiltonian dynamics, and sample the microcanonical ensemble.
In practical applications, alternative ensembles like canonical or isothermal-isobaric (NPT) ensembles are usually more desirable,
so various methods have been developed to sample the required ensemble by modifying the Hamiltonian dynamics.

One class of approaches to generate the desired ensemble is the extended
phase space methods. 
For example, the Nos\'e-Hoover~\cite{nose1984molecular,hoover1985canonical}, Nos\'e-Hoover chain~\cite{martyna1992nose} and stochastic Nos\'e-Hoover thermostats
are proposed to generate the canonical ensemble;
The Andersen~\cite{andersen1980molecular}, Parrinello-Rahman~\cite{parrinello1981polymorphic, parrinello1980crystal} and
Martyna-Tuckerman-Klein~\cite{martyna1994constant,martyna1996explicit} barostats are proposed to generate the NPT ensemble.
These methods share the idea of extending the physical phase-space (positions, velocities of particles and the simulation cell)
by extra variables that control the temperature and/or pressure of the system.
The dynamics of the extended system is carefully designed to fulfill the condition that
if the trajectory is ergodic, then
the marginal stationary distribution in the physical phase-space is consistent with that of the desired ensemble.

Langevin dynamics is an alternative method for generating the canonical ensemble.
It has been proved that the ergodicity is guaranteed under mild restrictions~\cite{mattingly2002ergodicity, mattingly2002geometric},
therefore the convergence to the canonical distribution is ensured in the limit of infinitely long simulation time.
Because of this advantage, Langevin dynamics has attracted increasing attention recently,
and various integration schemes were developed
in order to improve the accuracy of numerical simulations~\cite{bussi2007accurate,melchionna2007design,bou2010long,leimkuhler2013robust,leimkuhler2013rational,gronbech2013simple,peters2014stochastic}.
The first attempt of using the Langevin dynamics in generating the NPT ensemble was
from Feller et.al.~\cite{feller1995constant} and  Kolb and D\"unweg~\cite{kolb1999optimized}, who proposed the Langevin dynamics
for both the particle degrees of freedom and the volume of the simulation cell (isotropic cell fluctuation).
\redc{These work are recently improved by Gr\o nbech-Jensen and Farago~\cite{gronbech2014constant} and Di Pierro et.~al.~\cite{dipierro2015stochastic}.}
Quigley and Probert coupled the Parrinello-Rahman dynamics with Langevin stochastic terms
to extend the method to the fully flexible cell motions (anisotropic cell fluctuation)~\cite{quigley2004langevin,quigley2005constant}.
However, it is not possible, as the authors stated, to prove that the  stationary distribution of the dynamics is subject to the NPT ensemble
\redc{when the cell motion is stochastic {and} rotation-eliminated}.
Therefore, theoretically, the cell motions in this approach should be deterministic, and the convergence to the NPT distribution is not guaranteed, although the NPT distribution is one of the stationary distributions.
  In practice, the authors recommended to accept the cell stochastic motions at the cost of rigor in theory.

In this work, we propose a fully-flexible-cell NPT Langevin dynamics that allows explicit stochastic components in both the particle and cell motions,
due to which the convergence of configurational distribution to the NPT ensemble is naturally ensured in the infinitely long time limit.
We start by defining the Hamiltonian for an extended phase-space composed of scaled coordinates and simulation cell variables.
Then the Langevin dynamics of this system can be directly written down, and the Boltzmann stationary distribution is obtained in the extended phase-space.
In order to have a direct description of the particle motions in terms of physical coordinates,
the scaled coordinates are transformed back, and the Langevin dynamics is reformulated accordingly by using Ito's formula.
The configurational stationary distribution is then proved to be consistent with that of
the NPT ensemble by transforming the Boltzmann distribution back to the physical coordinates in the same way.
To develop the numerical scheme,
we start by considering the Fokker-Planck equation that is equivalent to the Langevin equation,
then the scheme is formulated by splitting 
the single-step propagator of the Fokker-Planck equation according to Trotter's theorem.
By construction, the proposed scheme is of second order accuracy with respect to the time-step size.

  Before moving to the main results of this work, 
  it should be noted that
  it is in general difficult to check if the convergence to the NPT distribution is achieved in a finite simulation time.
  One might check the convergence of some properties of the system, for example the free energy profile along a certain reaction coordinate~\cite{kelly2004calculation},
  but the choice of indicating properties depends on the nature of the system and what is wanted from the simulation,
  and they are usually not sufficient to prove the convergence of the distribution.
  Therefore, we do NOT intend to investigate the ergodicity or the speed of convergence to the desired ensemble distribution in numerical simulations.
  The significance of this work is to propose a new fully-flexible-cell NPT Langevin dynamics
  that takes the \emph{theoretical} advantage of ergodicity, and can be used as an alternative to the existing NPT simulation methods.

This paper is organized as follows:
The development of NPT Langevin dynamics is discussed in detail in Sec.~\ref{sec:tmp2}.
The discretization of Langevin dynamics is provided in Sec.~\ref{sec:tmp3}.
In Sec.~\ref{sec:tmp4}, we validate the NPT Langevin dynamics by a solid argon system of triclinic region cell.
This work is concluded in Sec.~\ref{sec:tmp5}.

\section{The Langevin equations of motion}
\label{sec:tmp2}

We denote the particle positions in the system by $\vect r_1, \cdots, \vect r_N$, where $N$ is the number of particles in the system.
The simulation cell matrix is denoted by $\vect h = [\vect h_1, \vect h_2, \vect h_3]$, where
$\vect h_\alpha, \alpha=1,2,3$ are cell vectors. The scaled (direct) coordinate of a particle $\vect s_i$
is defined by
$  \vect r_i = \vect h \vect s_i$.
In order to generate the fully-flexible-cell NPT ensemble, all components of the cell vectors
are allowed to fluctuate.
We define the kinetic energy of the system by
\begin{align}\label{eqn:k-energy}
  K = \sum_i \frac12 m_i (\vect h \dvs_i)^2 + \sum_{\tensorab}\frac12 M_{\tensorab} \dot h_{\tensorab}^2,
\end{align}
where $M_\tensorab$ is the fictitious mass corresponding to the motion of $h_\tensorab$ that is the $\beta$-th component of the $\alpha$-th cell vector.
The first term on the RHS of \eqref{eqn:k-energy} is different from the \emph{physical} kinetic energy of the system, which is~$\sum \frac12 m_i [ d(\vect h s_i)/dt ]^2$.
The consequence of this difference will be discussed in detail later.
It is worth noting that the Parrinello-Rahman barostat~\cite{parrinello1980crystal,parrinello1981polymorphic} also uses Eq.~\eqref{eqn:k-energy} as the definition of kinetic energy.
The Lagrangian of the system is defined as 
\begin{align}
  \mathcal L (\{\vect s_i\}, \{\dvs_i\}, \vect h, \dvh)
  = K  - (U + P\det(\vect h) + \redc{\chi \kt \ln[\det(h)]}),
\end{align}
where $U = U(\vect h\vect s_1, \cdots, \vect h\vect s_N)$ is the potential energy of the system, $P$ is the target pressure,
\redc{$\chi$ is a constant that depends only on the dimension of the system and will be discussed in Remark 1, $T$ is the target temperature and $k_B$ is the Boltzmann constant }.
The generalized momenta corresponding to $\vect r_i$ and $\vect h$ are
\begin{align}
  \mypi_i = m_i \vect h^\top\vect h \dvs_i, \quad p^h_\tensorab = M_\tensorab \dot h_\tensorab,
\end{align}
respectively.
Therefore, the Hamiltonian of the system yields
\begin{align}\label{eqn:hamiltonian}
  \haml (\{\vect s_i\}, \{\mypi_i\}, \vect h, \vect p_h)
  =
  \sum_{\tensorab} \frac{(p^h_\tensorab)^2}{2M_\tensorab} +  \sum_i \frac{(\vect h^{-\top}\mypi_i)^2}{2m_i}  + U + P\det (\vect h)  + \redc{ \chi \kt \ln[\det(h)]}.
\end{align}
The first term on the RHS of \eqref{eqn:hamiltonian} is the kinetic energy of cell vectors, and the summation of the last three terms gives the instantaneous enthalpy.
The Langevin dynamics is defined by
\vskip -.8cm
\begin{subequations}\label{eqn:lang-scaled}
\begin{align}\label{eqn:lang-scaled-0}
  \dot{\vect s}_i
  &=
  \frac{\partial \haml}{\partial \mypi_i} \\\label{eqn:lang-scaled-1}
  \dot{\mypi}_i
  &=
  -\frac{\partial \haml}{\partial \vect s_i} - \Gamma_i\frac{\partial \haml}{\partial\mypi_i} + \Sigma_i\dot{\vect W_i}\\\label{eqn:lang-scaled-2}
  \dot{h}_{\tensorab}
  &=
  \frac{\partial \haml}{\partial p^h_{\tensorab}}\\    \label{eqn:lang-scaled-3}
  \dot{p}^h_{\tensorab}
  &=    
  -\frac{\partial \haml}{\partial h_{\tensorab}}    
  -\hat\gamma_{\tensorab}\frac{\partial \haml}{\partial p^h_{\tensorab}} + \hat\sigma_{\tensorab} \dot W_\tensorab,
\end{align}  
\end{subequations}
where $W_i$, $W_\tensorab$ denote the standard Wiener processes, which are independent for different particle and cell degrees of freedom.
The friction $\Gamma_i$ and noise magnitude $\Sigma_i$ are $\vect h$ dependent matrices defined by
$\Gamma_i = \gamma m_i \vect h^\top\vect h$, $\Sigma_i = \sigma \sqrt{m_i}\, \vect h^\top$, where $\sigma^2 = 2\gamma\kt$,
then the fluctuation-dissipation theorem $\Sigma_i\Sigma_i^\top = 2\Gamma_i\kt$ holds for \eqref{eqn:lang-scaled-0}--\eqref{eqn:lang-scaled-1}.
We define $\hat\gamma_\tensorab = M_\tensorab\gamma_\tensorab$ and $\hat\sigma_\tensorab = \sqrt{M_\tensorab}\sigma_\tensorab$,
where $\sigma^2_\tensorab = 2 \gamma_\tensorab\kt$, then
the fluctuation-dissipation theorem $\hat\sigma_\tensorab^2 = 2\hat\gamma_\tensorab\kt$ holds for \eqref{eqn:lang-scaled-2}--\eqref{eqn:lang-scaled-3}.
It is well known that the stationary probability density of the Langevin dynamics~\eqref{eqn:lang-scaled} is
\begin{align}
  \rho_\equi \propto
  \exp\Big[ -\frac{1}{\kt} \haml (\{\vect s_i\}, \{\mypi_i\}, \vect h, \vect p^h)\Big].
\end{align}

It is more convenient to represent the Langevin dynamics \eqref{eqn:lang-scaled} 
in physical coordinates, so we introduce the following transformation
\begin{align}\label{eqn:var-trans}
  \vect r_i = \vect h\vect s_i, \quad \vect p_i = \vect h^{-\top}\mypi_i.
\end{align}
By using Ito's formula, and writing down all partial derivatives explicitly,
we reach
\begin{subequations}\label{eqn:aniso-npt-eom}
\begin{align}\label{eqn:aniso-npt-eom-0}
  \dvr_i &= \frac1{m_i}\vect p_i + \dvh\vect h^\inv\vect r_i \\\label{eqn:aniso-npt-eom-1}
  \dvp_i & = - \partial_i U - \vect h^{-\top}\dvh^\top\vect p_i -   \gamma \,\vect p_i +  \sqrt{m_i}\, \sigma\,\dot W_i\\\label{eqn:aniso-npt-eom-2}
  \dot h_\tensorab
  &=
  \frac1{M_\tensorab} p^h_\tensorab\\ \label{eqn:aniso-npt-eom-3}
  \dot{p}^h_\tensorab
  & =
  \det (\vect h) \Big[ \Big (P_\ins - P - \redc{ \frac{\chi\kt}{\det(h)} } \Big) \vect h^{-\top} \Big]_\tensorab
  - \gamma_\tensorab \, p^h_\tensorab
  + \sqrt{M_\tensorab}\,\sigma_\tensorab\, \dot W_\tensorab,
\end{align}
\end{subequations}
where $P_\ins$ is the instantaneous pressure tensor defined by
\begin{align}\label{eqn:press-ins}
  P_\ins  =
  \frac{1}{\det(\vect h)} \sum_i
  \Big(
  \frac1{m_i} \vect p_i\otimes\vect p_i + \vect F_i\otimes\vect r_i
  \Big).
\end{align}
It can be easily shown that the Jacobian determinant of the transform~\eqref{eqn:var-trans} is 1,
therefore,
by integrating out the cell vector momenta,
the equilibrium probability density generated by dynamics~\eqref{eqn:aniso-npt-eom-0}--\eqref{eqn:aniso-npt-eom-3}
is 
\begin{align}\label{eqn:invariant-meas}
  \rho_\equi 
  \propto
  \redc{  [\det(h)]^{-\chi}}
  \exp
  \Big[
  -\frac1\kt
  \Big(
  \sum_{i=1}^N \frac{\vect p_i^2}{2m_i}
  + U(\{\vect r_i\})
  + P\det(\vect h)
  \Big)
  \Big].
\end{align}

\noindent
\textbf{Remark 1}:
\redc{
  It was proposed by Ref.~\cite{martyna1994constant} that $\chi$ takes the value of $d-1$, where $d$ is  the dimension of the system.
  If the rotation of the simulation cell is eliminated by taking $h$ as an upper (or lower) triangular matrix, then $\chi$ should take $(d-1)/2$ (see Appendix~\ref{app:tmp2} for more details).
  The term $-\chi\kt/\det(h)$  on the RHS~of Eq.~\eqref{eqn:aniso-npt-eom-3} can be effectively treated as a correction to the pressure difference $P_\ins - P$, and 
  vanishes under the thermodynamic limit, viz.~the volume of the system $\det(h)$  goes to infinity.
}

\noindent
\textbf{Remark 2}:
It should be noted here that the particle momentum $\vect p_i$ is not exactly the physical momentum $\vect p^\phys_i$.
As a matter of fact, we have the relation:
\begin{align}
  \vect p^\phys_i =
  \vect p_i + {m_i}\dvh\vect h^\inv \vect r_i.
\end{align}
This inconsistency is the direct consequence of the definition of the kinetic energy~\eqref{eqn:k-energy}, in which the particle contribution is not the physical kinetic energy.
There is no substantial difficulty in using the physical definition, however, the derivation of the Langevin dynamics would become much more complicated.
In most applications, only the correctness of system configuration is of importance.
The current definition guarantees the configurational distribution to be consistent with that of the  NPT ensemble.
\redc{
  It is also worth noting that
  the particle momentum $p_i$ rather than $p^\phys_i$ should be used
  to estimate an equilibrium quantity that is an average over momentum-dependent instantaneous values (e.g.~temperature and pressure).
  By using $p_i$, the equilibrium distribution~\eqref{eqn:invariant-meas}  matches the form of the NPT distribution, so the average
  converges to the NPT equilibrium quantity under the infinity long time limit,
  and is independent with the parameters (frictions and the fictitious mass) in the Langevin dynamics.
}

\noindent
\textbf{Remark 3}:
The choice of the parameters in the Langevin dynamics~\eqref{eqn:aniso-npt-eom} has been extensively discussed in literature, e.g.~\cite{kolb1999optimized,quigley2004langevin}.
We pick up the friction coefficients based on a rule of thumb~\cite{quigley2004langevin}:
$\gamma = \gamma_\tensorab = \omega_T / 2\pi = 1/\tau_T$,
where $\omega_T$ denotes the frequency of the thermostat and $\tau_T = 2\pi/\omega_T$ is the time-scale of the thermostat.
In order to provide the way of choosing the fictitious mass $M_\tensorab$, we firstly assume that the cell matrix $h$ is diagonal (a cuboid cell).
With a first order expansion of the pressure with respect to the cell fluctuation in $h_{11}$, $h_{22}$ and $h_{33}$, we have
\begin{align}\label{eqn:diff-p-expan}
  P - P^0 = -\frac{1}{\kappa h^0_{11}h^0_{22}h^0_{33}} (h_{11}h_{22}h_{33} - h^0_{11}h^0_{22}h^0_{33}),
\end{align}
where $\kappa = -\frac1V \frac{\partial V}{\partial P}$ is the compressibility.
The superscript ``0'' denotes the equilibrium value of the corresponding variable. We further assume that the fluctuation from the equilibrium value is small,
i.e.~$\vert h_{\tensoraa} - h_{\tensoraa}^0\vert, \ \alpha=1,2,3$ are small.
Arranging the equation~\eqref{eqn:diff-p-expan} in the component-wise way and preserving only the first order fluctuations on the RHS, we have
\begin{multline*}
  \frac13\Big[
  (P_{11} - P_{11}^0) + (P_{22} - P_{22}^0) + (P_{33} - P_{33}^0)
  \Big] \\
  =
  -\frac{1}{\kappa h^0_{11}h^0_{22}h^0_{33}} \Big [
  (h_{11} - h^0_{11}) h^0_{22}h^0_{33} +
  (h_{22} - h^0_{22}) h^0_{11}h^0_{33} +
  (h_{33} - h^0_{33}) h^0_{11}h^0_{22} \Big].
\end{multline*}
One possible solution to the equation is
\begin{align}\label{eqn:diff-p-component}
  P_{\tensoraa} - P_{\tensoraa}^0  = - \frac3{\kappa h^0_{\tensoraa}} (h_{\tensoraa} - h^0_{\tensoraa}), \quad \alpha = 1,2,3.
\end{align}
By inserting \eqref{eqn:aniso-npt-eom-2} and \eqref{eqn:diff-p-component} into \eqref{eqn:aniso-npt-eom-3}, and discarding the friction and noise, \redc{$\chi$}, and higher order terms, we have
\begin{align*}
  M_\tensoraa\ddot h_\tensoraa
  = - \frac{3\det(h^0)}{\kappa  (h^0_\tensoraa)^2} (h_{\tensoraa} - h^0_{\tensoraa}),
\end{align*} 
which is the equation of motion of harmonic oscillator $h_{\tensoraa}$ with spring constant $k_\tensoraa = \frac{3\det(h^0)}{\kappa  (h^0_\tensoraa)^2}$ and equilibrium position $h_{\tensoraa}^0$.
By using the relation $\omega_\tensoraa = \sqrt{k_\tensoraa/M_\tensoraa}$, where $\omega_\tensoraa$ is the barostat frequency, we derive the expression for the fictitious mass
\begin{align}
  M_\tensoraa
  = \frac{3 \det(h^0)}{\kappa (h^0_\tensoraa)^2} \Big(\frac{1}{\omega_\tensoraa} \Big)^2
  = \frac{3 \det(h^0)}{\kappa (h^0_\tensoraa)^2} \Big(\frac{\tau_\tensoraa}{2\pi}  \Big)^2,
\end{align}
where $\tau_\tensoraa$ is the time-scale of the barostat.
For simplicity, we choose the off-diagonal value of fictitious mass by
\begin{align}
  M_\tensorab
  = \frac{3 \det(h^0)}{\kappa (h^0_\tensoraa)^2} \Big(\frac{\tau_\tensorab}{2\pi}  \Big)^2.
\end{align}
In practice, it is often impossible to predict the equilibrium cell matrix $h^0$ before performing the simulation,
however, if the initial condition does not deviate very far from equilibrium, we take the initial value of $h$ as a reasonable guess for $h^0$.
The compressibility $\kappa$ can either take an experimental value,
or be estimated from short testing simulations using the formula: $\kappa =  (\langle V^2\rangle - \langle V\rangle^2) / (\kt\langle V\rangle$).

It is worth noting that
regardless of the choice of the parameters,
the Langevin dynamics~\eqref{eqn:aniso-npt-eom} samples the configurational distribution of the NPT ensemble at infinitely long time limit.
The difference in using different parameters lies in the sampling efficiency.
If the time-scales of the thermo- and barostat were chosen too large,
then the temperature and pressure of the system would not be adjusted in a responsive way, and the sampling of the NPT ensemble would be too slow.
On the other hand, if they were chosen too small, then an unnecessarily small time-step would be needed to keep the MD simulation stable.
Therefore, it was suggested that the inverse time-scales be chosen just below the typical molecular frequency~\cite{kolb1999optimized}.

\section{Discretize the Langevin dynamics}
\label{sec:tmp3}

The evolution of a system governed by the Langevin dynamics \eqref{eqn:aniso-npt-eom-0} -- \eqref{eqn:aniso-npt-eom-3} is equivalently described by the following Fokker-Planck equation:
\begin{align}
  \frac{\partial \rho}{\partial t} = \opfwd \rho,
\end{align}
where $\rho(t, \{r_i\}, \{p_i\}, h, p^h) $ is the time dependent probability density defined on the phase space.
$\opfwd$ is the infinitesimal generator, which can be factorized as
\begin{align}
  \opfwd = \opfwd_K + \opfwd_U + \opfwd_{O} + \opfwd^h_K + \opfwd^h_U + \opfwd^h_{O},
\end{align}
with each term defined by
\begin{align*}
  \opfwd_K &=
  \sum_i\Big[
  \frac{\vect p_i}{m_i} + \dvh\vect h^\inv\vect r_i
  \Big] \cdot
  \frac{\partial}{\partial \vect r_i}\\
  \opfwd_U &=
  \sum_i\Big[
  -\partial_i U - \vect h^{-\top}\dvh^\top\vect p_i
  \Big] \cdot
  \frac{\partial}{\partial \vect p_i}\\
  \opfwd_{O} &=
  \sum_{i=1}^N
  \Big[\,
  3\gamma +
  \gamma \vect p_i \frac{\partial}{\partial \vect p_i}
  + \frac{m_i\sigma^2}{2}\frac{\partial^2}{\partial \vect p_i^2}    
  \,\Big] \\
  \opfwd^h_K &=
  \sum_\tensorab
  \frac{p^h_\tensorab}{M_\tensorab}
  \frac{\partial}{\partial h_\tensorab}  \\
  \opfwd^h_U &=
  \sum_\tensorab \Big\{ \det(\vect h) \Big [ \Big(\vect P_\ins - \vect P - \redc{ \frac{\chi\kt}{\det(h)} }  \Big ) \vect h^{-\top}\Big ]_\tensorab   \Big\}
  \frac{\partial}{\partial p^h_\tensorab}  \\
  \opfwd^h_{O} &=
  \sum_\tensorab
  \Big[\,
  \gamma_\tensorab +
  \gamma_\tensorab p^h_\tensorab \frac{\partial}{\partial p^h_\tensorab}
  + \frac{M_\tensorab\sigma_\tensorab^2}{2}\frac{\partial^2}{\partial (p^h_\tensorab)^2}    
  \,\Big].
\end{align*}

Given this factorization, the single-step propagator $e^{\Delta t\opfwd}$ ($\Delta t$ being the time-step)
can be split by the Trotter theorem:
\begin{align}\label{eqn:trotter-split}
  e^{\Delta t \opfwd} =
  e^{\frac {\Delta t}2 \opfwd^h_U }
  e^{\frac {\Delta t}2 \opfwd_U }
  e^{\frac {\Delta t}2 \opfwd^h_K }
  e^{\frac {\Delta t}2 \opfwd_K }
  e^{ {\Delta t} \opfwd^h_{O} }
  e^{ {\Delta t} \opfwd_{O} }
  e^{\frac {\Delta t}2 \opfwd_K }
  e^{\frac {\Delta t}2 \opfwd^h_K }
  e^{\frac {\Delta t}2 \opfwd_U }    
  e^{\frac {\Delta t}2 \opfwd^h_U }
  + \mathcal O(\Delta t^3).
\end{align}
This style of splitting is actually the ``BAOAB'' scheme proposed by Ref.~\cite{leimkuhler2013robust}.
The authors argued that the BAOAB splitting is more accurate than other schemes in the sense of configurational sampling.
The action of propagator $e^{ {\Delta t} \opfwd_K }$ and $e^{ {\Delta t} \opfwd_U }$ corresponds to evolve $\vect r_i$ and $\vect p_i$ by $\Delta t$ under the ordinary differential equation
$\dvr_i = \frac1{m_i}\vect p_i + \dvh\vect h^\inv\vect r_i $ and 
$\dvp_i = - \partial_i U - \vect h^{-\top}\dvh^\top\vect p_i$, respectively.
Here we adopt the convention that the cell matrix $\vect h$ is an upper triangular matrix, then
one has to solve, in general,
$\dot {\vect x} = \vect b + \vect A\vect x $
with $\vect A$ being an upper or lower triangular matrix. In the current work, the analytic solution
of
$\dot {\vect x} = \vect b + \vect A\vect x $
in the upper and lower triangular cases
are denoted by $\vect x(t) = \vect S_u( \vect x(0), t, \vect b, \vect A)$ and $\vect x(t) = \vect S_l(\vect x(0),t, \vect b, \vect A)$, respectively.
The explicit forms of function $\vect S_u$ and $\vect S_l$ are provided in Appendix~\ref{app:solve-s}.
The action of propagator $e^{ {\Delta t} \opfwd_{O} }$ and $e^{ {\Delta t} \opfwd^h_{O} }$ corresponds to
evolve variables $\vect p_i$ and $p^h_\tensorab$  by $\Delta t$ under the Ornstein-Uhlenbeck process.
In general, an Ornstein-Uhlenbeck process $dp = -\gamma pdt + \sigma\sqrt{m} dw_t$
can be explicitly solved by $p(t) = e^{-\gamma t} p(0) + \frac{\sigma}{\sqrt{2\gamma}}\sqrt{1- e^{-2\gamma t}}\sqrt{m} R$,
where $R$ is a random number subject to the normal distribution with vanishing mean and unit variance, i.e.~$\mathcal N(0,1)$.

Discarding the higher order terms and applying from left to right the propagators on the RHS of \eqref{eqn:trotter-split} yields the
following numerical scheme:
\begin{algorithmic}[1]
  \While{MD continues}
  \State $p^h_\tensorab  \gets p^h_\tensorab + {\Delta t}/ 2\, \det(\vect h) [(\vect P_\ins - \vect P - {\chi\kt}/{\det(h)}) \vect h^{-\top}]_\tensorab  $
  \State $\vect p_i  \gets \vect S_l (\vect p_i, {\Delta t}/2, -\partial_i U,  - \vect h^{-\top}\dvh^\top)$
  \State $h_\tensorab    \gets h_\tensorab   + {\Delta t}/ 2\, ({p^h_\tensorab}/{M_\tensorab}$)
  \State $\vect r_i  \gets \vect S_u (\vect r_i, {\Delta t}/2, {\vect p_i}/{m_i},  \dvh\vect h^\inv)$
  \State $p^h_\tensorab
  \gets e^{-\gamma_\tensorab {\Delta t}} p^h_\tensorab + \sqrt {1 - e^{-2\gamma_\tensorab \Delta t}} \:\sqrt{M_\tensorab\kt}\, R$
  \State $\vect p_i
  \gets e^{-\gamma \Delta t} \vect p_i + \sqrt {1 - e^{-2\gamma \Delta t}} \: \sqrt{m_i\kt} \vect R$
  \State $\vect r_i  \gets \vect S_u (\vect r_i, {\Delta t}/2, {\vect p_i}/{m_i},  \dvh\vect h^\inv)$
  \State $h_\tensorab    \gets h_\tensorab   + {\Delta t}/2\, ( {p^h_\tensorab}/{M_\tensorab}$)
  \State Compute the force for each particle
  \State $\vect p_i  \gets \vect S_l (\vect p_i, {\Delta t}/2, -\partial_i U,  - \vect h^{-\top}\dvh^\top)$
  \State Compute the instantaneous pressure tensor 
  \State $p^h_\tensorab  \gets p^h_\tensorab + {\Delta t}/ 2\, \det(\vect h) [(\vect P_\ins - \vect P - {\chi\kt}/{\det(h)}) \vect h^{-\top}]_\tensorab  $
  \EndWhile
\end{algorithmic}  
The operators $\opfwd_O$ and $\opfwd_O^h$ are mutable, therefore, lines 6 and 7 in the algorithm can be swapped.

\section{Numerical results}
\label{sec:tmp4}

The NPT Langevin method was implemented in the in-house molecular dynamics simulation package MOASP
developed on infrastructure JASMIN~\cite{mo2010jasmin}.
In order to validate the theory and the implementation, we tested a solid argon system modeled by the Lennard-Jones interaction:
\begin{align}
  U(r) = \frac{C_{12}}{r^{12}} - \frac{C_{6}}{r^{6}} + C,
\end{align}
where $r$ is the distance between a pair of particles. $C_{12} =
2.71507\times 10^{-7}$~$\mathrm{kJ\:mol}^{-1}\mathrm{nm}^{-12}$ and $C_6 =
1.72685\times 10^{-4}$~$\mathrm{kJ\:mol}^{-1}\mathrm{nm}^{-6}$ are parameters 
taken from the CHARMM27 force field~\cite{foloppe2000all,mackerell2000all}.
$C$ is a shifting constant that ensures the continuity of energy at the cut-off.
The cut-off radius was chosen to
be 0.9~nm in all simulations.
The neighbor list was build for particles that are at most 1.1~nm apart, and was updated every 20 time-steps.
If not stated otherwise, the time-step of integration was chosen to be 0.001~ps.

\begin{figure}
  \centering
  \includegraphics[width=0.48\textwidth]{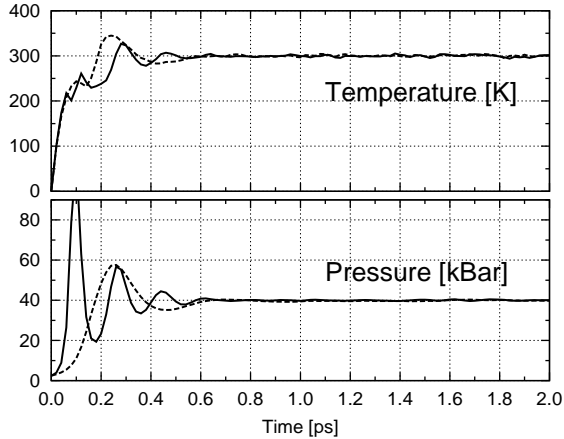}
  \caption{The equilibration of the Lennard-Jones argon system toward 300~K and 40~kBar.
    The
    simulation starts from a perfect FCC configuration at 0~K, and lasts
    for 100~ps.  The plots present the equilibration of temperature
    (upper plot) and pressure (lower plot) at the first 2~ps.
    The solid line uses a compressibility of $4.5\times10^{-5}$~$\mathrm{Bar}^{-1}$,
    and the dashed line uses a compressibility of $0.8\times10^{-5}$~$\mathrm{Bar}^{-1}$.
  }
  \label{fig:equi}
\end{figure} 

An initial configuration of perfect FCC crystal was prepared
by extending the Bravais lattice cell of $\vert \vect h_1\vert = \vert\vect h_2 \vert = \vert\vect h_3\vert = 1.825$~nm and $\alpha = \beta = \gamma = 60^\circ$
by $30\times20\times 20$ times along three cell vectors, respectively.
Therefore, the system contained 12,000 atoms in total.
An 100~ps equilibration simulation that used this configuration and zero initial velocities was conducted at 300~K and 40~kBar.
The initial guess of the compressibility was $4.5\times10^{-5}$~$\mathrm{Bar}^{-1}$ (which was actually a value taken from the liquid water under ambient condition).
The time-scales of thermostat and barostat were set to 0.1~ps and 0.5~ps, respectively.
The system was successfully equilibrated to the desired thermodynamic state within only 1~ps (the solid lines in Fig.~\ref{fig:equi}), and 
the initial FCC solid structure was stable under this thermodynamic condition.
The finial coordinates and velocities of atoms were recorded for productive simulations.
The compressibility, $0.8\times10^{-5}$~$\mathrm{Bar}^{-1}$, was estimated from this simulation, and was used for all following simulations.
Since the initial guess of the compressibility was much larger than $0.8\times10^{-5}$~$\mathrm{Bar}^{-1}$,
the speed of equilibration was actually faster than the user specified thermo- and barostat time-scales.
We conducted the equilibration again with the correct compressibility,
and found that the speed of equilibration was roughly the same as the specified thermo- and barostat time-scales (the dashed lines in Fig.~\ref{fig:equi}).

\begin{figure}
  \centering
  \includegraphics[width=0.45\textwidth]{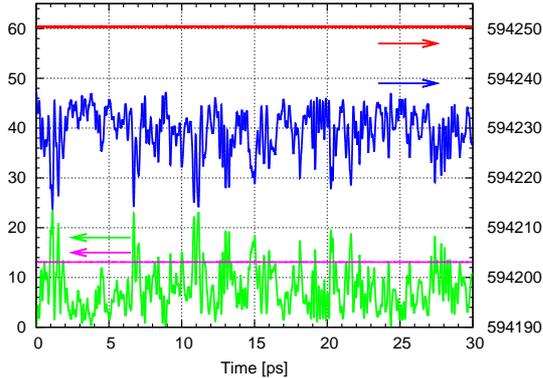}
  \caption{The conservation of Hamiltonian when the friction and noise vanish.
    The time evolution of the cell kinetic energy (green), \redc{$\chi\kt\ln[\det(h)]$ (pink)}, the instantaneous enthalpy (blue) and the Hamiltonian (red)
    are presented. The  kinetic energy and \redc{$\chi\kt\ln[\det(h)]$} use the left y-axis, while the instantaneous enthalpy and the Hamiltonian
    use the right y-axis, as the arrows in the Figure indicate. The unit of the energy is kJ/mol.
  }
  \label{fig:consv-hamil}
\end{figure} 

An effective way to validate the correctness of the equations and software implementation is to check the conservation of Hamiltonian~\eqref{eqn:hamiltonian} when
the friction and noise in \eqref{eqn:aniso-npt-eom} vanish.
We performed this simulation with the initial positions and velocities from the previous equilibration,
and plot the evolution of the cell kinetic energy, the instantaneous enthalpy, \redc{$\chi\kt\ln[\det(h)]$}  and the Hamiltonian in  Figure~\ref{fig:consv-hamil}.
A perfect conservation of the Hamiltonian is observed.
It is worth mentioning that 
the cell kinetic energy and the instantaneous enthalpy
fluctuate at the magnitude of roughly $\pm 5$~kJ/mol (see Fig.~\ref{fig:consv-hamil}),
and that the kinetic and potential energy of particles fluctuate at the magnitude of roughly $\pm 300$~kJ/mol (not shown).

\begin{figure}
  \centering
  \includegraphics[width=0.23\textwidth]{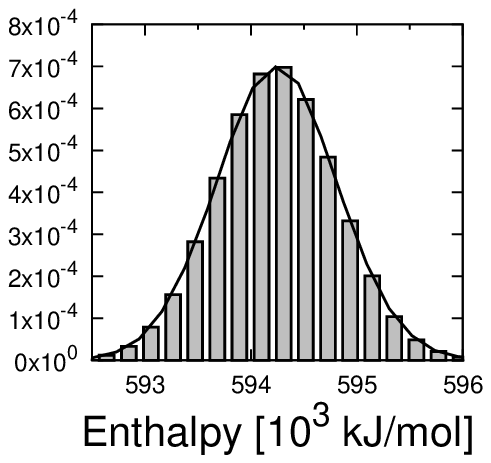}
  \includegraphics[width=0.23\textwidth]{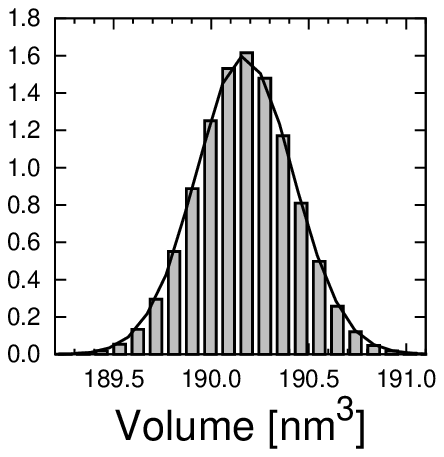}
  \caption{The distribution of the instantaneous enthalpy (left) and cell volume (right) of the solid argon system.
    The gray bars present the probability densities calculated from
    the 10,000~ps Langevin NPT simulation (this work)
    at 300~K and 40~kBar. The solid lines present the same
    probability densities calculated from the reference simulation (see the text for more details).
  }
  \label{fig:tmp1}
\end{figure} 

\begin{figure}
  \centering
  \includegraphics[width=0.23\textwidth]{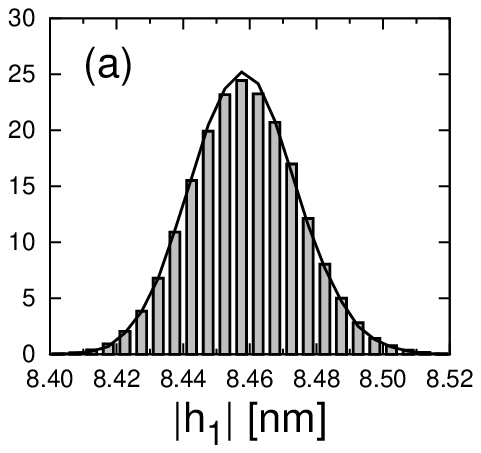}
  \includegraphics[width=0.23\textwidth]{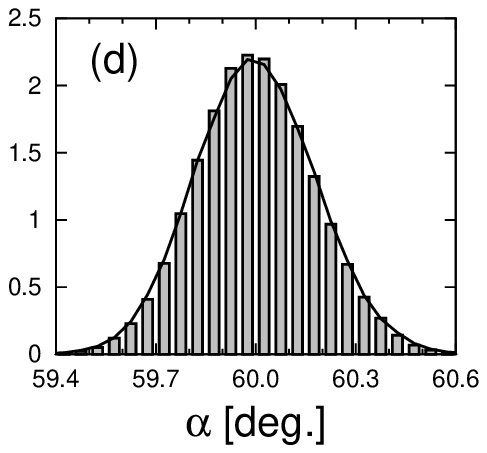}\\
  \includegraphics[width=0.23\textwidth]{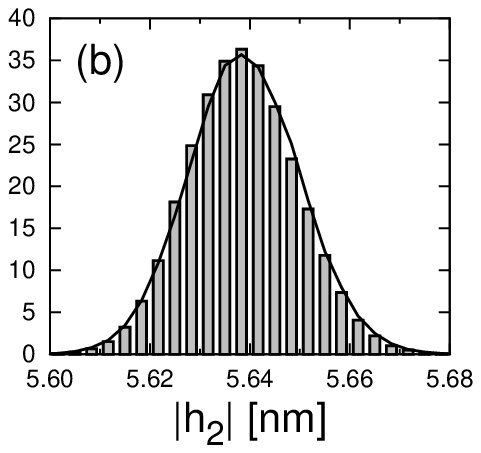}
  \includegraphics[width=0.23\textwidth]{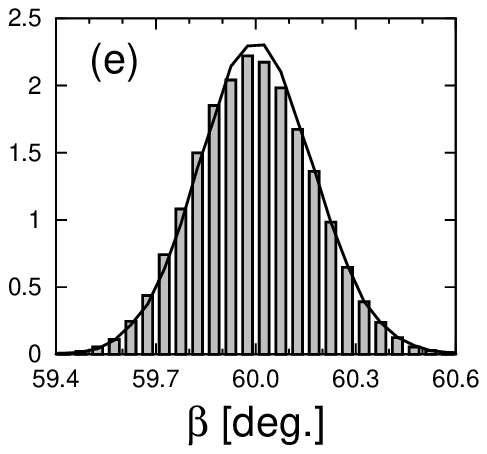}\\
  \includegraphics[width=0.23\textwidth]{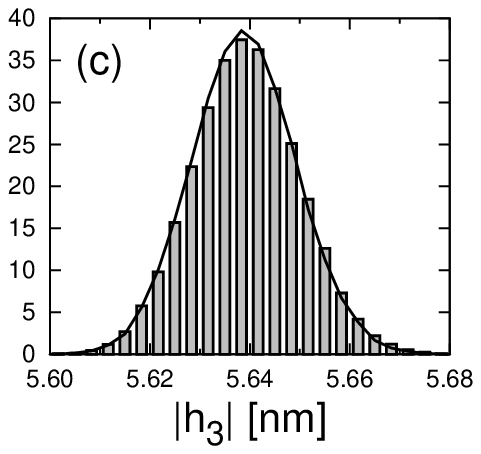}
  \includegraphics[width=0.23\textwidth]{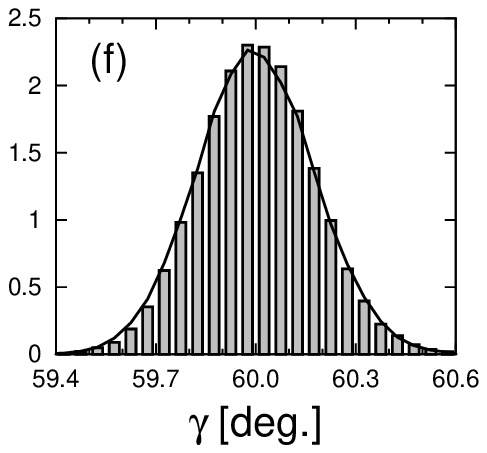}
  \caption{The distribution of cell vectors.  
    Plot (a) -- (c) on the left column present
    the probability densities of lengths of three cell vectors, i.e.~$\vert \vect h_1\vert $, $\vert \vect h_2\vert$ and $\vert \vect h_3\vert$, respectively.
    Plot (d) -- (f) on the right column  present
    the probability densities of three angles $\alpha$, $\beta$ and $\gamma$ between them, respectively.
    ($\alpha$ is the angle between $\vect h_2$ and $\vect h_3$.
    $\beta$ is the angle between $\vect h_1$ and $\vect h_3$.
    $\gamma$ is the angle between $\vect h_1$ and $\vect h_2$.)
    The simulated system is the same as Fig.~\ref{fig:tmp1}.
  }
  \label{fig:tmp2}
\end{figure}

\begin{figure}
  \centering
  \includegraphics[width=0.46\textwidth]{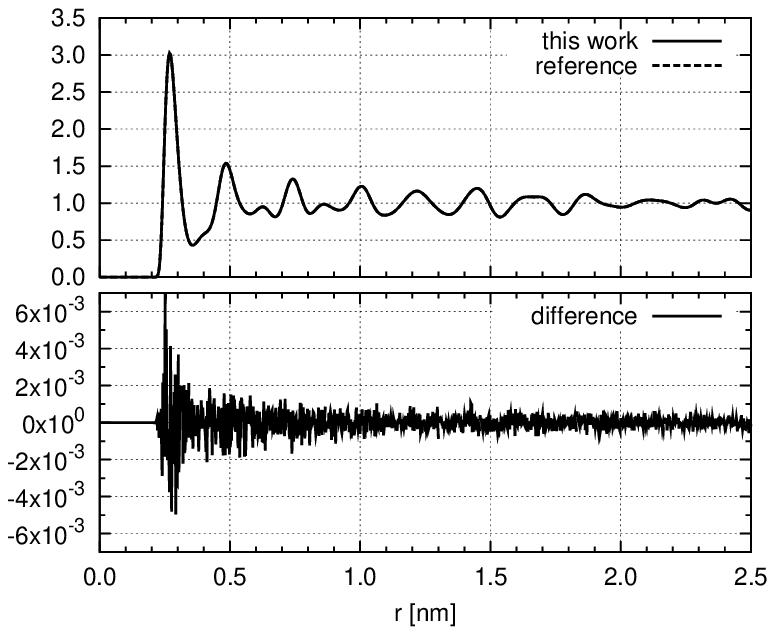}  
  \caption{The radial distribution functions calculated by the NPT Langevin (this work) and reference simulations.
    The two RDFs are plotted together in the upper plot;
    The difference between them is presented in the lower plot.
  }
  \label{fig:rdf}
\end{figure}

The productive NPT simulation lasted for 10,000~ps, and the data
collection started from 100~ps. The instantaneous enthalpy and cell vectors were recorded every 0.02~ps. 
We plot the distribution of the instantaneous enthalpy
and cell volume in Fig.~\ref{fig:tmp1}.
The distribution of the lengths of cell vectors and the angles between them
are presented in Fig.~\ref{fig:tmp2}.
In the figures, the results of the NPT Langevin dynamics are compared and consistent with those of 
a reference simulation that uses velocity-rescaling thermostat~\cite{bussi2007canonical} and Parrinello-Rahman barostat~\cite{parrinello1980crystal,parrinello1981polymorphic}.
The initial condition and the other parameters were set to be the same as the Langevin simulation.
The reference simulation is considered to be reliable, because it was conducted by a well-tested MD simulation package Gromacs~\cite{hess2008gromacs,pronk2013gromacs} (version 4.6.5).
The radial distribution functions (RDFs) calculated by Langevin and the reference simulation
are shown to be overlapping in the upper plot of Fig.~\ref{fig:rdf}. From the lower plot of Fig.~\ref{fig:rdf},
we observe that the difference between them is dominated by the statistical uncertainty.
This means that the solid structures sampled by Langevin dynamics reproduce those of the reference simulation.

\begin{figure}
  \centering
  \includegraphics[width=0.46\textwidth]{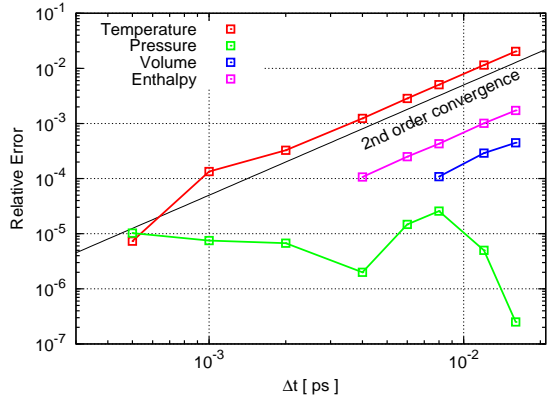}  
  \caption{
    The convergence of temperature, pressure, volume and enthalpy
    with respect to the time-step ($\Delta t$) of the integration.
    The $x$-axis is the time-step in unit of picosecond, and the $y$-axis is the relative error.
    The reference values of the volume and enthalpy are obtain by the simulation using time-step 0.0005~ps.
    The relative errors of volume below $\Delta t$ = 0.008~ps and enthalpy below $\Delta t$ = 0.004~ps are dominated by statistical uncertainties, so they are not presented for clarity.
  }
  \label{fig:converge-dt}
\end{figure}

\redc{
  The NPT Langevin dynamics is numerically solved by the  integration scheme in Sec.~\ref{sec:tmp3}
  that inevitably introduces error by time-discretization.
  To investigate its effect on temperature, pressure, volume and enthalpy computations, we present
  the relative errors vs. time-step size in a log-scaled plot (Fig.~\ref{fig:converge-dt}).
  The reference values of the volume and enthalpy are obtained by a simulation using the time-step of 0.0005~ps. 
  The statistical uncertainty in estimating the relative errors are,
  on average, $4.7\times 10^{-5}$, $3.3\times 10^{-5}$, $1.1\times10^{-4}$ and $1.0\times10^{-4}$ (95\% confidence level)
  for temperature, pressure, volume and enthalpy, respectively.  
  The numerical results prove that the integration scheme is of second-order accuracy in computing the temperature, volume and enthalpy.
  The accuracy of pressure computation is remarkably high, so the errors are dominated by the statistical uncertainty ($3.3\times10^{-5}$), and do not present the dependency on the time-step size.
  The relative errors in computing the cell vector lengths and angles are also dominated by the statistical uncertainty, and are not presented here.
}

\section{Conclusions and discussions}
\label{sec:tmp5}

In this work, we proposed a new fully-flexible-cell Langevin dynamics in NPT ensemble.
Our approach couples  stochastic terms to both the particle and cell degrees of freedom,
and is proved to correctly sample the configurational stationary distribution of the NPT ensemble.
We noted that the choice of the working parameters (friction coefficients and the fictitious masses of cell vectors) 
does not affect the sampling of the NPT ensemble in the infinitely long time limit,
however it determines the  sampling efficiency and the size of the MD time-step.
Therefore, we suggested a practical guide that automatically computes the fictitious mass and the friction coefficients by using the user specified thermo-/barostat time-scales,
and the compressibility of the simulated material.
In order to solve the Langevin equations by computers,
a discretization scheme was developed by using the Trotter splitting of the single-step propagator of the Fokker-Planck equation.
This scheme is, by construction, of second order accuracy.
A solid argon system modeled by the Lennard-Jones interaction was simulated to validate the proposed Langevin dynamics and the numerical scheme.
The conservation of the Hamiltonian in the case of vanishing friction and noise was firstly check, and the correctness of the equations and the software implementation was confirmed.
Then the equilibrium distributions of enthalpy, cell volume and vectors were calculated, and were found to be in good consistency
with those  calculated by a reference simulation (conducted by package Gromacs~4.6.5).
The accuracy of the equilibrium structure of the system was verified
by comparing the radial distribution function with the reference simulation.

\redc{
There are, roughly speaking,
two classes of approaches for sampling canonical or isothermal-isobaric ensembles:
The first is the deterministic approaches utilizing the extended phase space dynamics,
while the second is the stochastic approaches introducing random noise in the dynamics.
One can also find methods that combine the ideas from both classes of approaches, e.g.~Ref.~\cite{samoletov2007thermostats,leimkuhler2009gentle,dipierro2015stochastic}.
The main concern about the deterministic approaches is the ergodicity~\cite{hoover1985canonical},
the theoretical proof of which  is, as far as we see, still unavailable.
Thus the advantage of the stochastic approaches is the theoretically guaranteed ergodicity, 
which indicates that one can expect converged sampling of the desired ensemble in the infinitely long time limit.
However, it should be noted that the theoretical proof of ergodicity does not necessarily ensure better performance in practice,
because no real simulation is infinitely long.
Both classes of approaches face the issue of how efficiently the phase space is explored.
As discussed in Sec.~\ref{sec:intro},
this challenging question can be numerically investigated by checking the convergence of indicating properties on a case-by-case basis.
As a matter of fact,
one can find examples in the literature showing that the stochastic approaches are more precise in computing some properties (e.g.~\cite{kelly2004calculation}),
while other examples indicating that the deterministic approaches are more efficient (e.g.~\cite{omelyan2013generalised}).
Since it is impossible to derive a conclusive answer on which class of approaches is superior,
we suggest to provide the users both options in MD simulation packages.
}

\redc{
In this work, the numerical scheme for solving the Langevin equations is designed by a second order Trotter splitting of the single-step propagator.
It worth mentioning that the proposed scheme is not the only possibility to achieve the time-discretization.
Alternative schemes can be designed, for example, by manipulating the order of operators in Eq.~\eqref{eqn:trotter-split}~\cite{leimkuhler2013robust},
or by minimizing the truncation errors of extended high-order splitting schemes with free parameters~\cite{omelyan2002optimized,omelyan2002optimized2,omelyan2003symplectic}.
The studies along this direction will be carried out in the future.
}

Although the Langevin dynamics is an efficient sampling tool of the equilibrium ensembles,
the computed dynamical properties like time correlation functions are usually wrong,
because the stochastic terms in the momentum equations break the physical dynamics of the system
that is described by the Newton's equations of motion.
Therefore, in the cases where the dynamical properties are intended to be computed precisely, more carefully designed methods, e.g.~local Langevin thermostat, should be used~\cite{wang2015adaptive}.

\section{Acknowledgment}
The authors acknowledge the valuable discussions with Aiqing Zhang, Xu Liu and Jianjun Liu, and the technical supports from them.
H.W. thanks the fruitful discussions with Dr.~Wei Zhang from Freie Universit\"at Berlin.
The authors gratefully acknowledge the financial support from
National High Technology Research and Development Program of China under Grant 2015AA01A304.
X.G. is supported by the National Science Foundation of China under Grants 91430218 and 61300012.
H.W. is supported by the National Science Foundation of China under Grants 11501039 and 91530322.

\appendix
\section{The value of constant $\chi$}
\label{app:tmp2}
\redc{
  In this section we follow the arguments of Ref.~\cite{martyna1994constant}.
  To separate the volume fluctuation from the total fluctuation of $h$,
we let $h = V^{1/d} h_0$, where $V$ is the volume of the system, and $h_0$ with $\det(h_0)= 1$ accounts for the shape of the simulation cell.
The partition function of the NPT ensemble (isotropic pressure control) reads
\begin{align}\label{eqn:partition}
  \Delta = \int dV d h_0 \, e^{-{PV}/\kt} Q(V,h_0)\, \delta (\det(h_0) - 1)
\end{align}
where $Q$ is the canonical partition function.
When transforming $h_0$ back to $h$ in the integration, the measure transform $dh = (V^{1/d})^{d^2} dh_0$ is used for case that the $d^2$ components of the tensor $h$ are not constrained.
If the rotation of $h_0$ (or equivalently $h$) is eliminated 
by taking $h_0$ as an upper triangular matrix ($(h_0)_\tensorab = 0,\ \alpha < \beta$),
the number of degrees of freedoms in the tensor $h_0$ is $d(d+1)/2$, therefore
we have the measure transform $dh = (V^{1/d})^{d(d+1)/2} dh_0$.
In general, we denote $dh = (V^{1/d})^{\nu} dh_0$, and integrate the partition function Eq.~\eqref{eqn:partition} over $V$:
\begin{align}\nonumber
  \Delta & = \int dV d h \, V^{-\nu/d} \, e^{-{PV}/\kt} Q(h)\, V \delta (\det(h) - V)\\
  & = \int d h \, [\det(h)]^{-(\nu/d-1)} e^{-{P\det(h)}/\kt} Q(h).
\end{align}
This means the constant $\chi$ should take the value of $\nu/d-1$.
When the simulation cell is allowed to rotate, $\nu = d^2$, so $\chi$ takes $d-1$.
When the rotation of the simulation cell is removed, $\nu = d(d+1)/2$, so $\chi$ takes $(d-1)/2$.
}

\section{Solve the equation $\dot{ \vect x} = \vect b + \vect A\vect x$}
\label{app:solve-s}

The ordinary differential equation  $\dot{ \vect x} = \vect b + \vect A\vect x$ can be solved
analytically. We start with the case that $\vect A$ is an upper triangular matrix:
\begin{align}\label{eqn:dxbAu}
  \left [
    \begin{matrix}
      \dot x_0\\
      \dot x_1\\
      \dot x_2
    \end{matrix}
  \right ]
  =
  \left [
    \begin{matrix}
      b_0\\
      b_1\\
      b_2
    \end{matrix}
  \right ]
  +
  \left [
    \begin{matrix}
      a_{00} & a_{01} & a_{02}\\
      0 & a_{11} & a_{12}\\
      0 & 0 & a_{22}\\
    \end{matrix}
  \right ]
  \cdot
  \left [
    \begin{matrix}
      x_0\\
      x_1\\
      x_2
    \end{matrix}
  \right ].
\end{align}
The solution is given by:
\begin{align} \nonumber
  x_0(t) = \,
  &   x_0(0) e^{a_{00}t} + t\, b_0 F_1(0, a_{00}t) \\ \nonumber
  & + t\, a_{01} x_1(0) F_1(a_{00}t, a_{11}t) + t^2\, a_{01} b_1 F_2 (a_{00}t, 0, a_{11}t)  \\ \nonumber
  & + t\, a_{02} x_2(0) F_1(a_{00}t, a_{22}t) + t^2\, a_{02} b_2 F_2 (a_{00}t, 0, a_{22}t)  \\ 
  & + t^2 a_{01}a_{12} x_2(0) F_2 (a_{00}t, a_{11}t, a_{22}t) + t^3 a_{01}a_{12} b_2 F_3 (a_{00}t, a_{11}t, 0, a_{22}t)\\ \nonumber
  x_1(t) =\,
  &   x_1(0) e^{a_{11}t} + t\, b_1 F_1(0, a_{11}t) \\ 
  & + t\, a_{12} x_2(0) F_1(a_{11}t, a_{22}t) + t^2\, a_{12} b_2 F_2 (a_{11}t, 0, a_{22}t)\\
  x_2(t) = \,
  & x_2(0) e^{a_{22}t} + t\, b_2F_1 (0, a_{22}t),
\end{align}
where the function $F_1$, $F_2$ and $F_3$ are defined to be
\begin{align}
  F_1 (A,B)     &= \frac{e^A - e^B}{A-B}\\
  F_2 (A,B,C)   &= \frac{1}{B-C} ( F_1(A,B) - F_1(A,C) )\\
  F_3 (A,B,C,D) &= \frac{1}{C-D} ( F_2(A,B,C) - F_2(A,B,D) ).
\end{align}
It should be noticed that the definitions have singularity when any two of the $A$, $B$, $C$ and $D$ are equal,
and we do not exclude these cases in real simulations. 
Noticing that we only need the solution of \eqref{eqn:dxbAu} at small $t$,
the singularity can be avoided by Taylor expansion of the exponential functions around 0.
Following this idea, we reach the serial expansions of $F_1$, $F_2$ and $F_3$:
\begin{align}
  F_1 (A,B)           &= \sum_{k=0}^\infty \frac1{(k+1)!} \sum_{\substack{0\leq\alpha,\beta\leq k\\\alpha+\beta = k }} A^\alpha B^\beta\\
  F_2 (A,B,C)         &= \sum_{k=0}^\infty \frac1{(k+2)!} \sum_{\substack{0\leq\alpha,\beta,\gamma\leq k\\\alpha+\beta+\gamma = k}} A^\alpha B^\beta \\
  F_3 (A,B,C,D)       &= \sum_{k=0}^\infty \frac1{(k+3)!} \sum_{\substack{0\leq\alpha,\beta,\gamma,\delta\leq k\\\alpha+\beta+\gamma+\delta = k}} A^\alpha B^\beta C^\gamma D^\delta.
\end{align}

For the case that $\vect A$ is a lower triangular matrix: 
\begin{align}\label{eqn:dxbAl}
  \left [
    \begin{matrix}
      \dot x_0\\
      \dot x_1\\
      \dot x_2
    \end{matrix}
  \right ]
  =
  \left [
    \begin{matrix}
      b_0\\
      b_1\\
      b_2
    \end{matrix}
  \right ]
  +
  \left [
    \begin{matrix}
      a_{00} & 0     & 0    \\
      a_{10} & a_{11} & 0    \\
      a_{20} & a_{21} & a_{22}\\
    \end{matrix}
  \right ]
  \cdot
  \left [
    \begin{matrix}
      x_0\\
      x_1\\
      x_2
    \end{matrix}
  \right ],
\end{align}
the solution can be written down in a similar way:
\begin{align} 
  x_0(t) = \,
  & x_0(0) e^{a_{00}t} + t\, b_0F_1 (0, a_{00}t) \\ \nonumber
  x_1(t) =\,
  &   x_1(0) e^{a_{11}t} + t\, b_1 F_1(0, a_{11}t) \\ 
  & + t\, a_{10} x_0(0) F_1(a_{11}t, a_{00}t) + t^2\, a_{10} b_0 F_2 (a_{11}t, 0, a_{00}t)\\\nonumber
  x_2(t) = \,
  &   x_2(0) e^{a_{22}t} + t\, b_2 F_1(0, a_{22}t) \\ \nonumber
  & + t\, a_{21} x_1(0) F_1(a_{22}t, a_{11}t) + t^2\, a_{21} b_1 F_2 (a_{22}t, 0, a_{11}t)  \\ \nonumber
  & + t\, a_{20} x_0(0) F_1(a_{22}t, a_{00}t) + t^2\, a_{20} b_0 F_2 (a_{22}t, 0, a_{00}t)  \\ 
  & + t^2 a_{21}a_{10} x_0(0) F_2 (a_{22}t, a_{11}t, a_{00}t) + t^3 a_{21}a_{10} b_0 F_3 (a_{22}t, a_{11}t, 0, a_{00}t).
\end{align}


\end{document}